\scrollmode
\documentclass[12pt]{amsart}
\usepackage{graphicx, verbatim}
\title[Generalized simultaneous conjugacy problem and the Algebraic Eraser]{On
the cryptanalysis of the generalized simultaneous conjugacy search problem and the
security of the Algebraic Eraser}

\newif\ifsubmission
\submissiontrue
\submissionfalse
\ifsubmission
\else
\author{Paul E. Gunnells}
\address{Dept. of Mathematics and Statistics\\
UMass Amherst\\
Amherst, MA 01003}
\email{gunnells@math.umass.edu}
\thanks{This research was partially supported by SecureRF Corporation}
\fi

\date{28 February 2011}

\newcommand{\Z}{\mathbb{Z}}
\newcommand{\alf}[1]{|#1|_{a}}
\newcommand{\x}{\mathbf{x}}
\newcommand{\y}{\mathbf{y}}
\DeclareMathOperator{\Tlnf}{Tlnf}
\begin{document}

\maketitle
\begin{abstract}
The \emph{Algebraic Eraser} (AE) is a cryptographic primitive that can
be used to obscure information in certain algebraic cryptosystems.
The \emph{Colored Burau Key Agreement Protocol} (CBKAP), which is
built on the AE, was introduced by I.~Anshel, M.~Anshel, D.~Goldfeld,
and S.~Lemieux \cite{aagl} in 2006 as a protocol suitable for use on
platforms with constrained computational resources, such as RFID and
wireless sensors.  In 2009 A.~Myasnikov and A.~Ushnakov proposed an
attack on CBKAP \cite{attack} that attempts to defeat the
\emph{generalized simultaneous conjugacy search problem}, which is the
public-key computational problem underlying CBKAP.  In this paper we
investigate the effectiveness of this attack.  Our findings are that
success of the attack only comes from applying it to short keys, and
that with appropriate keys the attack fails in 100\% of cases and does
not pose a threat against CBKAP.  Moreover, the attack in
\cite{attack} makes assumptions about CBKAP that do not hold in
practical implementations, and thus does not represent a threat to the
use of CBKAP in applications.
\end{abstract}

\section{Introduction}

\subsection{}
In \cite{aagl} I.~Anshel, M.~Anshel, D.~Goldfeld, and S.~Lemieux
propose a key agreement protocol intended for use on low-cost
platforms with constrained computational resources.  Such platforms
typically arise in radio frequency identification (RFID) networks and
wireless sensor networks.  The protocol is built on the
\emph{Algebraic Eraser} (AE), a cryptographic primitive that disguises
information in many algebraic cryptosystems, such as those built on
conjugation problems in braid groups.  For more details, including a
formal description of the AE, we refer to \cite[\S 2]{aagl}.

The security of the AE is based on the hardness of the
\emph{generalized simultaneous conjugacy search problem} (GSCSP) which
can be described as follows.  Suppose $G$ is a group and $(X)$ is a
property potentially satisfied by elements of $G$ (i.e., elements
satisfy property $(X)$ if and only if they satisfy certain identities
in $G$).  Then given $y_{1},\dotsc ,y_{n}\in G$, the associated GSCSP
is to find elements $z, a_{1},\dotsc ,a_{n},$ such that $y_{i} =
za_{i}z^{-1}$ for all $i$ and the $a_{i}$ satisfy $(X)$.  Note that
GSCSP is a broader problem than the simultaneous conjugacy search
problem where the elements $a_1, \ldots, a_n$ are known and there is
no specified property $(X)$.

As an example implementation of a protocol based on the AE, the
authors of \cite{aagl} present the \emph{Colored Burau Key Agreement
Protocol} (CBKAP).  The algebraic structure underlying this protocol
is the braid group $B_{n}$, and an essential part of CBKAP is a
trusted third party (TTP) algorithm that chooses secret data in
$B_{n}$.  We give this data in detail in \S \ref{s:setup}, and for now
only mention that the TTP chooses a secret element $z\in B_{n}$---the
\emph{conjugator}---and uses it to produce finite lists of conjugates
$\{V_{i} \}, \{W_{i} \}\subset B_{n}$.  These sets are made available
for the protocol's users.  As shown in \cite[\S 6]{aagl}, knowledge of
$z$ allows one to break CBKAP.  If both sets $\{V_{i} \}, \{W_{i} \}$
are published, the security of CBKAP relies on the assumed difficulty
of recovering $z$ from these sets. This is an instance of the GSCSP.

\subsection{}
In \cite{attack} A.~Myasnikov and A.~Ushnakov present an attack on
CBKAP that relies on both sets of conjugates $\{V_{i} \}, \{W_{i} \}$
being known.  Instead of trying to determine $z$, they try to find an
alternative element $\zeta$ that can play the role of $z$ in the
attack in \cite[\S 6]{aagl}.  They also heuristically analyze the
difficulty of finding $\zeta$, and make the claim that they can
recover the secret conjugator in all instantiations of the TTP at the
security levels proposed in \cite{aagl}.

\subsection{}
In this paper, we report on tests we performed with the attack in
\cite{attack}.  We tested the attack on randomly generated TTP data at
a variety of security parameters.  We also tested some of the
heuristic assumptions in \cite{attack} that form the core of the
attack.

We found that for suitable choices of the parameters, the attack fails
in 100\% of cases.  More precisely, for low TTP data length, the
attack in \cite{attack} is indeed successful in recovering $z$, and
thus in breaking CBKAP.  However, as lengths increase, the attack
becomes far less successful and eventually fails in 100\% of all
cases.  We also found that some of the heuristics underlying their
attack are too optimistic when word lengths become long, as one would
find in a typical deployment of CBKAP in a constrained computational
setting. 

Our tests suggest that the apparent power of the attack in
\cite{attack} comes from using it against poorly chosen TTP data, in
particular against braid words that are short.  Moreover, with
appropriate TTP data the attack poses no threat against CBKAP, even
for data leading to small public/private key sizes that may be
successfully deployed in low cost platforms with constrained
computational resources.  

\subsection{} Finally, we also remark that \cite{attack} uses heavily
the assumption that both sets $\{V_{i} \}, \{W_{i} \}$ are known to
the attacker.  Indeed, this assumption can be found in \cite{aagl}.
However, in most practical implementations of CBKAP this will not be
the case. For example, see \cite{aggeg} where it is shown that the AE
version of the El Gamal public key encryption algorithm \cite{elgamal}
requires only one of the sets $\{V_{i} \}, \{W_{i} \}$ to be made
public. In this case, the attack in \cite{attack} cannot even be
applied, and thus fails completely.

\section{The TTP algorithm and the attack}\label{s:setup}

\subsection{}
Let $B_{n}$ be the braid group on $n$ strands.  We denote the Artin
generators by $s_{1},\dots ,s_{n-1}$; they satisfy the defining
relations $s_{i}s_{i+1}s_{i}=s_{i+1}s_{i}s_{i+1}$ and $s_{i}s_{j} =
s_{j}s_{i}$ if $|i-j|>1$.  Let $\Delta$ be the half-twist $(s_{1}\dots
s_{n-1}) (s_{1}\dots s_{n-2})\dots (s_{1}s_{2})s_{1}$, whose square
generates the center of $B_{n}$.

To set up an instance of CBKAP, the TTP performs the following algorithm:
\begin{enumerate}
\item Choose a freely reduced word $z$ in the generators $s_{i}$ and
their inverses.
\item Choose two subgroups $B_{A}, B_{B} \subset B_{n}$ that are
mutually commuting: $ab = ba$ for all $a\in B_{A}, b\in B_{B}$.
\item Choose $2N$ words $v_{1},\dots ,v_{N}\in B_{A}$ and $w_{1},\dots
,w_{N}\in B_{B}$, and form the conjugates $zv_{1}z^{-1}, \dots , zv_Nz^{-1}, \; z w_1 z^{-1}, \ldots,
zw_{N}z^{-1}$.
\item For $i=1,\dotsc ,N$:
\begin{enumerate}
\item compute the left normal form \cite{wp} of $v_{i}$ and reduce the
result modulo $\Delta^{2}$;\label{reducev}
\item let $V_{i}$ be a braid word corresponding to the element
obtained in (\ref{reducev});
\item compute the left normal form of $w_{i}$ and reduce the result
modulo $\Delta^{2}$;\label{reducew}
\item let $W_{i}$ be a braid word corresponding to the element
obtained in (\ref{reducew}).
\end{enumerate}
\end{enumerate}

The lists $\{V_{i} \}$ and $\{W_{i} \}$ are made available to the
implementers of CBKAP, and the element $z$ is kept secret.  The
fundamental computational problem to break the protocol is the
following: \emph{Given the lists $\{V_{i} \}, \{W_{i} \}$ of disguised
(rewritten using the braid relations) conjugates, find $z$ and the
original words $\{v_{i}\}, \{w_{i}\}$.}  If one knows $z$, then an
attack on CBKAP was already described in \cite[\S 6]{aagl}.

\subsection{} Now we turn to the attack in \cite{attack}, which begins
with the following observation.  To break CBKAP using the strategy in
\cite[\S 6]{aagl}, it is not necessary to know the original words
$v_{1},\dotsc ,w_{N}$, which were chosen from a specific pair of
mutually commuting subgroups of $B_{n}$.  In fact, to apply \cite[\S
6]{aagl} one only needs to find \emph{some} way to produce conjugates
of the published lists that lie in mutually commuting subgroups.  This
leads to the following computational problem, which the authors of
\cite{attack} call the \emph{simultaneous conjugacy separation search
problem} (SCSSP): \emph{Given the published lists $\{V_{i} \}, \{W_{i}
\}$, find an element $\zeta $ and integers $p_{1},\dots
,p_{N},q_{1},\dots ,q_{N}$ such that the two sets $\{w_{i}'\} =
\{\Delta^{2p_{i}} \zeta^{-1}W_{i}\zeta \mid i=1,\dots , N\}$ and
$\{v_{i}'\} = \{\Delta^{2q_{i}}\zeta^{-1}V_{i}\zeta \mid i=1,\dots ,
N\}$ are subsets of mutually commuting subgroups of $B_{n}$}.  The
element $\zeta$ is then applied in the linear attack described in
\cite[\S 6]{aagl}, in which it plays the role of the conjugator $z$.
Of course the original $z$ and exponents of $\Delta^{2}$ used in the
normal form reduction in steps (\ref{reducev}), (\ref{reducew}) will
solve the SCSSP, but there could be other choices that work as well.

Thus the attack falls naturally into two steps:
\begin{enumerate}
\item Determine the
exponents $p_{i}, q_{i}$.
\item Determine the conjugator
$\zeta$. 
\end{enumerate}
Both steps rely heavily on a function $|\cdot |_{a}\colon
B_{n}\rightarrow \Z$, the \emph{approximate length function}.  This
function, originally defined in \cite{length}, serves as a replacement
for the geodesic length $l\colon B_{n}\rightarrow \Z$ in the Cayley
graph of $B_{n}$.  We discuss this function more below, and for now
explain how it is used in attack.

\subsection{}
We begin with step (1).  Let $X$ be any element from the published
lists of disguised conjugates.  We want to find the associated
exponent $p$ of $\Delta^{2}$ that should be applied with $X$ to solve
the SCSSP.  Consider the set of integers
\begin{equation}\label{eq:lengths}
\{\alf{\Delta^{2j}X} \mid j\in \Z \}.
\end{equation}
We assume that the set \eqref{eq:lengths} attains a minimum at some
integer $p$.  This is our desired exponent for $X$.  We repeat the
procedure for all $V_{i}$ and $W_{i}$.  

\subsection{}
After finding all the exponents $p_{1},\dotsc ,q_{N}$, the next step
(2) is finding $\zeta$. To explain how this is done, we need more
notation.  Let $\mathbf{x} = (x_{1},\dotsc , x_{N})$ be a tuple of
words in $B_{n}$.  Let $|\mathbf{x}|_{a} = \sum |x_{i}|_{a}$ be the
total approximate length of $\x$.  For any $w\in B_{n}$ let
$\mathbf{x}^{w} = (w^{-1}x_{1}w, \dotsc , w^{-1}x_{N}w)$.

Now suppose we have two tuples $\x , \y$ that we know a priori can
be conjugated into two commuting subgroups.  Put $\zeta = 1$.  We
consider simultaneous conjugation of $\x ,\y $ by generators, and how
the total approximate length of the tuples $\x ,\y $ change.  In other
words, for each $\sigma \in \{s_{1}^{\pm 1}, \dotsc , s_{n-1}^{\pm 1}
\}$, let $\delta_{\sigma}$ be defined by
\[
\delta_{\sigma} = \alf{\x^{\sigma}} + \alf{\y^{\sigma}} - (\alf{\x} + \alf{\y}).
\]
If $\delta_{\sigma} > 0$, then conjugation by $\sigma$ makes the
tuples $\x , \y$ longer overall, and so $\sigma$ should not appear on
the end of a reduced expression for $\zeta$.  On the other hand, if
$\delta_{\sigma}<0$, then conjugation by $\sigma$
represents progress towards constructing $\zeta$.  We replace $\zeta$
with $\zeta \sigma$, replace $\x , \y$ with $\x^{\sigma},
\y^{\sigma}$, and repeat the process if $\x^{\sigma},
\y^{\sigma}$ are not supported on mutually commuting subgroups.  A
variation of this procedure keeps track of the sequence
$\sigma_{1},\sigma_{2},\dotsc$ and uses backtracking to try more
possibilities for $\zeta$.  

\section{The approximate length function}\label{s:length}

\subsection{}
A key role in the attack is played by the approximate length function
$\alf{\cdot}$, originally defined in \cite{length}.  To explain it we
need more notation.

Let $w\in B_{n}$ be represented by a reduced expression
$s_{i_{1}}\dotsb s_{i_{r}}$.  The \emph{main generator} of $w$ in this
expression is the generator $s_{j}$ such that $j$ is the minimal
subscript $i_{k}$ appearing in the expression.  A word is
\emph{Dehornoy reduced} if its main generator does not appear
simultaneously with its inverse \cite{dehornoy}.  Typically there are
many Dehornoy reduced expressions representing $w$, but one can write
a deterministic program to produce a unique one.  Following
\cite{dehornoy}, one can further use the reduction procedure to
produce a \emph{fully reduced word}.  Such a word is also Dehornoy
reduced, but satisfies additional properties that tend to make it
substantially shorter than the original word.  We assume this has been
done, and let $D (w)$ be the full reduction of $w$.

\subsection{}
The idea behind computing $|w|_{a}$ is to produce a word $w'$
equivalent to $w$ using a combination of full reduction and right
conjugation by $\Delta$.  The latter affects a reduced expression $w =
s_{i_1}^{\varepsilon_{i_1}}\cdots s_{i_k}^{\varepsilon_{i_k}}$ by
replacing each generator $s_{j}$ by its ``complement'' $s_{n-j}$:
\[
w^{\Delta} := \Delta^{-1}w\Delta = s_{n-i_1}^{\varepsilon_{i_1}}\cdots
s_{n-i_k}^{\varepsilon_{i_k}}.
\]

The algorithm to compute $\alf{w}$ works as follows.  We begin by
putting $w_{0} = w$ and $i=0$.  Let the word length of $z$ be denoted
$|z|$.  Then we apply the sequence
\begin{enumerate}
\item Increment $i$ and put $w_{i} = D (w_{i-1})$.\label{one}
\item If $|w_{i}|<|w_{i-1}|$, then
\begin{enumerate}
\item put $w_{i} = w_{i}^{\Delta}$ and
\item goto Step \ref{one}.
\end{enumerate}
\item Otherwise,
\begin{enumerate}
\item if $i$ is even output $w'=w_{i+1}^{\Delta}$;
\item if $i$ is odd output $w'=w_{i+1}$.
\end{enumerate}
\end{enumerate}
The output is a word $w'$ equivalent to $w$ with $|w'|\leq |w|$.
Finally we define $|w|_{a}$ to be $|w'|$.  In practice, for instance
as implemented in \cite{crag}, one does not repeat (1)--(2) until
$|w_{i}|\geq |w_{i-1}|$, but instead iterates a fixed number of times.

The authors of \cite{attack} claim that $\alf{\cdot}$ approximates the
geodesic length $l$ well enough so that two key properties hold.
First, they claim that for generic tuples $\x$ and words $w$, we have
$\alf{\x^{w}} > \alf{\x}$.  Next, they claim that $\alf{\cdot}$
approximately satisfies the triangle inequality.  Namely, we have
$\alf{w} + \alf{u} \geq \alf{wu}$ for generic words $w,u$.  Both
properties play a key role in the heuristic justifying the attack on
the TTP algorithm.

\section{Tests and findings}

\subsection{}
Our tests naturally split into two topics.  First we tested features
of the approximate length function, in particular how well the
computation of $\alf{\cdot}$ shortens words compared to full
reduction, as well as how well the approximate length function
satisfies the triangle inequality.  Next we tested the attack against
the TTP algorithm for a variety of randomly generated TTP data.

All algebraic computations with braid groups---including randomly
generating braid words, the implementation of the attack in
\cite{attack}, and the computation of the approximate length function
$\alf{\cdot}$---were performed using the C++ library \texttt{crag}
written by the authors of \cite{attack}, and distributed through the
Algebraic Cryptography Group at the Stevens Institute of Technology.
The code is freely available on the internet \cite{crag}.

\subsection{Approximate length function: reduction}\label{ss:alf} 
\begin{figure}[htb]
\begin{center}
\includegraphics{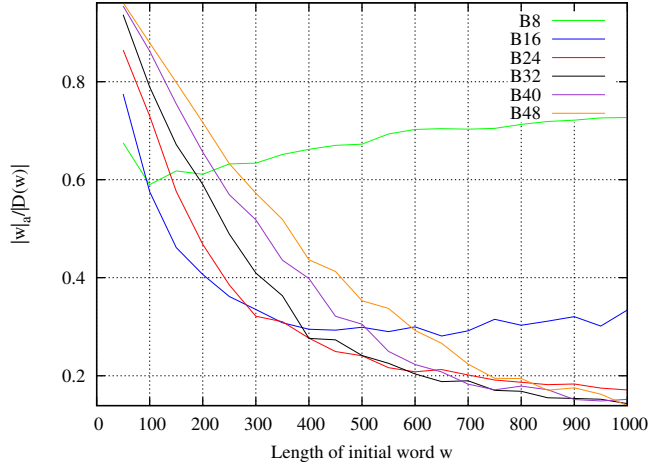}
\end{center}
\caption{Approximate length compared with length of full
reduction\label{fig:length} in various braid groups.}
\end{figure}

In
these tests we fixed a braid group $B_{n}$, then generated many freely
reduced words $w$ of various lengths and computed $\alf{w}/|D (w)|$.
The results for the groups $B_{8}, B_{16}, \dotsc , B_{48}$ are
plotted in Figure \ref{fig:length}; each data point represents 100
trials.

We found that when lengths of random generated initial words $w$ are
short relative to the rank $n$, the function $|w|_{a}$ is essentially
the length $|D (w)|$ of the full reduction $D (w)$ of $w$.  On the
other hand, when the length of the initial word increases, the ratio
$\alf{w}/|D (w)|$ drops quickly, even more rapidly as the number of
strands is increased.  As the length increases even more, it appears
that the ratio $\alf{w}/|D (w)|$ stabilizes to a constant.  The data
for $B_{8}$ may suggest that this constant is asymptotically $1$.  Thus it
appears that full reduction combined with conjugation by $\Delta$
can be used to produce rather short words, at least for a certain
range of initial lengths depending on the index.

We also checked how well $\alf{\cdot}$ performed before and after
applying Thurston left normal form $\Tlnf$ to a long freely reduced
word.  This normal form, described in \cite{wp}, can be used to prove
automaticity of the braid group.  For a randomly chosen freely reduced
word $w$ representing an element of the braid group, the word
$\Tlnf(w)$ is generally much longer than $w$.  We found that the
approximate length function is relatively insensitive to passing
through $\Tlnf$, and in particular $\alf{w}$ is very close to
$\alf{\Tlnf (w)}$ in most cases.  

\subsection{Approximate length function: triangle
inequality}\label{ss:ti} 
\begin{figure}[htb]
\begin{center}
\includegraphics{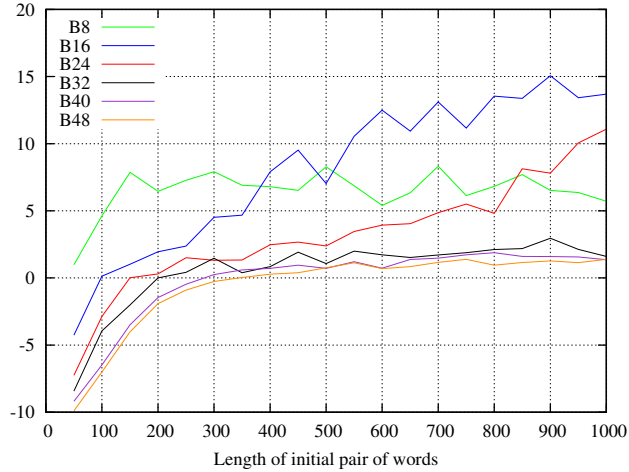}
\end{center}
\caption{Failure of the triangle inequality\label{fig:triangle} for
$\alf{\cdot}$ in various braid groups.}
\end{figure}

Next we checked how well the approximate
length function satisfies the triangle inequality $\alf{x}+\alf{y}\geq
\alf{xy}$.  We considered the same sequence of braid groups as in \S
\ref{ss:alf}.  After fixing $B_{n}$, we generated many freely reduced
words $x$, $y$ of the same length, and then computed the relative
error $100\cdot (\alf{xy}- (\alf{x}+\alf{y}))/\alf{xy}$.  The results
are shown in Figure \ref{fig:triangle}. Again each data point shows
the average over $100$ trials; the horizontal axis represents the
length of the randomly chosen $x$, $y$.

Thus Figure \ref{fig:triangle} shows the average relative error
between $\alf{xy}$ and $\alf{x}+\alf{y}$.  If this quantity is
negative, then the inequality holds on average, and if positive, then
$\alf{x}+\alf{y} < \alf{xy}$ on average.  The data indicates that for
short words, if the rank $n$ is increased then the triangle inequality
seems to hold, with $\alf{xy}$ considerably smaller than
$\alf{x}+\alf{y}$ on average.  But if the lengths of $x,y$ are
increased, then for all ranks ultimately the triangle inequality fails
to hold for $\alf{\cdot}$ on average.  The asymptotic behavior is not
clear.  We remark that the of course the triangle inequality holds in
all cases for the geodesic metric in the Cayley graph of $B_{n}$.

\subsection{The attack I: sensitivity to overall
length of TTP data}\label{ss:attack}

Next we tested the attack against randomly generated TTP data.  We
fixed the braid group $B_{16}$.  As above each data
point represents $100$ trials with given parameter choices.

First we ran a series of tests in which $N$ varied from $2$ to $10$;
recall that the TTP data consists of $2N$ conjugates divided into
two sets of $N$.  In these tests the elements $z$ and $v_{i}, w_{j}$
were chosen to have approximately the same word length.  The results
are plotted in Figure \ref{fig:random}.  The data clearly shows that
for small elements, i.e.~when $|z|, |v_{i}|, |w_{j}| \approx 67$ and
thus $|zv_{i}z^{-1}|, |zw^{j}z^{-1}| \approx 200$, the attack is
successful in almost all cases, regardless of the number of
conjugates.  But as the lengths are increased, the success rate drops
off quickly until the attack fails in all cases.  Moreover, the
success rate drops off more quickly as the number of conjugates is
increased.

\subsection{The attack II: dependence on relative sizes of $z$ and
$v_{i}$, $w_{j}$}\label{} 
\begin{figure}[htb]
\begin{center}
\includegraphics{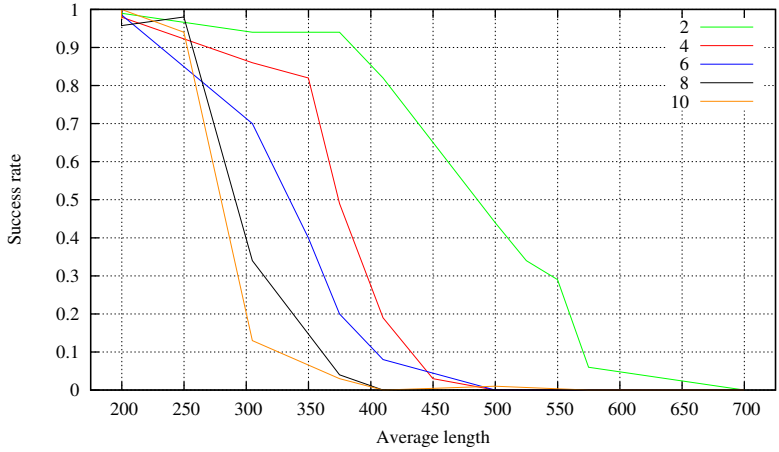}
\end{center}
\caption{\label{fig:random} Performance of the attack against randomly
generated TTP data.  The ambient braid group is $B_{16}$ and we show
data for different $N$.  \emph{Average length} refers to word lengths
of $\{zv_{i}z^{-1}, zw^{j}z^{-1}\}$. In this data the length of $z$ is
roughly equal to that of $v_{i}, w_{j}$.  In all cases the attack is
successful for short lengths and experiences a phase transition
to failure as lengths are increased.  The rapidity of the transition
depends on how many conjugates are used in the two sets.}
\end{figure}

Next we ran a series of tests to investigate
the performance of the attack when the lengths of $\{zv_{i}z^{-1},
zw_{j}z^{-1}\}$ are fixed and approximately equal, but $|z|$ is very
different from $|v_{i}|, |w_{j}|$.  We fixed $N$ to be $8$ and
$|zv_{i}z^{-1}| \approx |zw^{j}z^{-1}| \approx 350$, and considered
$|z| = 25, 50, \dotsc , 150$.  These parameters were chosen because
Figure \ref{fig:random} shows that the attack is successful about 15\%
of the time when the lengths of $z$, $v_{i}$, and $w_{j}$ are roughly
equal and the length of the conjugates is about $350$.  Hence at these
lengths one can evaluate the performance of the attack when
the relative lengths are varied.  

\begin{figure}[htb]
\begin{center}
\includegraphics{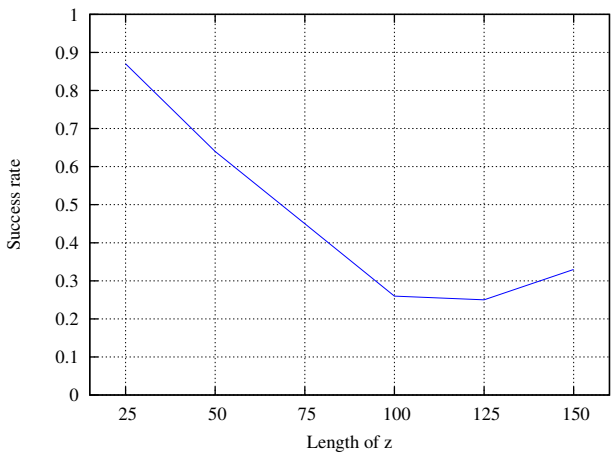}
\end{center}
\caption{\label{fig:zplusa} Investigating attack performance on
short words when the relative lengths of $z$ and $v_{i}, w_{j}$ are
varied.  For this data we work in in $B_{16}$ with $N=8$ and 
take $2|z|+|v_{i}|, 2|z|+|w_{j}| \approx 350$, in $B_{16}$ with $N=8$.
This graph refines one data point in Figure \ref{fig:random} at which
the attack is successful about 15\% of the time.}
\end{figure}

The results, shown in Figure \ref{fig:zplusa}, indicate that
increasing the length of $z$ significantly hampers the success of the
attack.  Comparison of Figure \ref{fig:zplusa} with the relevant data
point in the plot for $N=8$ in Figure \ref{fig:random} is also
instructive.  In the former, we have $|z|\approx 125, |v_{i}|, |w_{j}|
\approx 100$ with the total length of each conjugate about $350$.  In
the latter when $|z|\approx 125$ we have $|v_{i}|, |w_{j}| \approx
125$, so that the total length is $375$.  Thus the conjugators have
the same length and conjugates are slightly longer, yet the success
rate of the attack in the latter case is substantially lower.

\section{Conclusions and discussion}

\subsection{}
First, the approximate length function described in \cite{length} uses
a combination of full reduction and conjugation by $\Delta$ to produce
short expressions for words.  It does appear to offer an improvement
over full reduction, in that in almost all cases we tested it appears
to produce rather short words.  We conclude that this reduction
technique can be used to produce shorter words than those from full
reduction, at least for a set of lengths depending on the index.

\subsection{}
Next, the assumption that the triangle inequality holds for the
approximate length function $\alf{\cdot}$ appears to be too
optimistic.  As the lengths of words increase, the triangle inequality
apparently holds less and less often.  The asymptotic behavior is not
clear from our experiments, but nevertheless some of the data
($B_{16}, B_{24}$) suggests that the inequality might fail quite badly
in the long run.

\subsection{}
Regarding the attack on CBKAP described in \S \ref{ss:attack}, we find
that it is successful if the words $\{V_{i} \}, \{W_{i} \}$ are short,
and that the claims in \cite{attack} about the data they tested appear
valid.  However, as the lengths of $\{V_{i} \}, \{W_{i} \}$ increase,
the attack quickly loses power, and soon fails in all instances.
Furthermore, the attack does not seem robust against easily
implemented defenses.  Increasing the number of conjugates, for
instance, causes the attack to fail at much shorter word lengths.
Modifying key selection by varying the length of the conjugator also
adversely affects the attack's success.  Experiments also show that
selecting keys more carefully---for instance, but applying criteria to
randomly generated TTP data that go beyond length considerations
alone---also quickly hampers the performance of the attack.  We
conclude that the success of the attack seems mainly to be due to it
being applied to short words.

\bibliographystyle{amsplain_initials} \bibliography{paper}

\end{document}